\documentclass{article}
	
	\usepackage{arxiv}
	
	\usepackage{amsfonts}
	\usepackage{graphicx}
	\usepackage{epstopdf}
	\DeclareGraphicsExtensions{.eps}
	\usepackage{array}
	\newcolumntype{P}[1]{>{\centering\arraybackslash}p{#1}}
	\newcolumntype{M}[1]{>{\centering\arraybackslash}m{#1}}
	
	\newtheorem{prop}{Proposition}

	\usepackage{amsmath}
	\usepackage{multirow}
	\usepackage[noend]{algpseudocode}
	
	\def\v{{\mathbf v}}
	\def\V{{\mathbf V}}
	\def\S{{\mathbf S}}
	\def\C{{\mathbf C}}
	
	\title{Generalized Multi-Order Total Variation for Signal Restoration}
	
	\author{
		Sanjay~Viswanath\\
		Imaging Systems Lab\\
		Department of Electrical Engineering\\
		Indian Institute of Science\\
		Bangalore, Karnataka, India 560012 \\
		\texttt{sanjayv@iisc.ac.in} \\
		\And
		Muthuvel~Arigovindan\\
		Imaging Systems Lab\\
		Department of Electrical Engineering\\
		Indian Institute of Science\\
		Bangalore, Karnataka, India 560012 \\
		\texttt{mvel@iisc.ac.in} \\
	}

\begin{document}
	\maketitle
	
	\begin{abstract}
	Total Variation (TV)  based regularization has been widely applied in
	restoration problems due to its simple derivative filters based formulation 
	and robust performance. While first order TV suffers from staircase effect, 
	second order TV promotes piece-wise linear reconstructions. Generalized
	 Multi-Order Total Variation (GMO-TV) is proposed as a novel regularization
	  method which incorporates a new multivariate Laplacian prior on signal 
	  derivatives in a non-quadratic regularization functional, that utilizes subtle 
	  inter-relationship between  multiple order derivatives. We also propose a 
	  computational framework to automatically determine the weight parameters
	  associated with these derivative orders, rather than treating them as user
	   parameters. Using simulation results on ECG and EEG signals, we show 
	   that GMO-TV performs better than related regularization functionals.
	\end{abstract}
	

\section{Introduction}

Derivative-based regularization  approach has proven to be powerful   for signal restoration.
The required signal is computed as a minimizer of cost that is a weighted sum of
the goodness of fit to the measured data and a roughness measure, which is also
known as the regularization functional.  If the roughness functional is constructed as
the square of derivative values summed over the entire signal support,  the minimization
is equivalent to Tikhonov filtering. If the roughness functional is constructed as
the sum of absolute value of derivatives,  the functionals are called 
 Total Variation (TV) \cite{Rudin_tv1} regularization functionals.  These type of
 functionals and the 
 related extensions have been widely applied in signal/image restoration problems 
\cite{TV_ECGSmooth, TV_FetalECG, TV_Chambolle04, TV_Chan_BlindDecon, 
TV_CMDecon, TV_LiBlindDecon} due to their simple filtering based formulation and
 robust performance in the presence of noisy measurements. TV regularization 
 not only performs better than Tikhonov regularization \cite{tikh_1} in terms of preserving
 the resolution, but also retains the benefits of filter based formulation, which supports 
 parallelization \cite{TV_GpuPock, TV_GpuCT} and matrix free implementation with 
 reduced storage and computational requirements. 
	
Though first order TV \cite{Rudin_tv1} suffers from staircase artifacts \cite{TV_Ring, 
Combined_order_TV}, higher order methods \cite{Scherzer_tv2_98, 
TV_CombinedLysaker, Bergounioux2010, Bredies_TGV_2010, TGV2} yield better 
performance at the cost of increased computation. While second order TV 
\cite{Scherzer_tv2_98} deals with the application of second order derivatives in TV 
functional, Lysaker et al. \cite{TV_CombinedLysaker} included both first and second
 order derivatives in regularization using a parameterized function, where the 
 parameter was adjusted manually to change contribution from different derivatives for 
 increased restoration performance. Total Generalized Variation (TGV) 
 \cite{Bredies_TGV_2010} generalized the concept of bounded variation to arbitrary 
 orders and its second order form \cite{TGV2} has been applied with state of art 
 performance in restoration problems. The significant advantage with second order 
 TGV (TGV2) is its coupling of first and second order terms through an auxiliary 
 variable, which allows adaptive switching between derivative orders based on their
corresponding derivative values. Generalized Total Variation (GTV) \cite{GTV} is a 
recent extension that has been proposed to exploit the idea of group sparsity in 
second order signal derivatives, rather than direct sparsity of derivatives used in TV
functional. The GTV functional performs better than both first and second order TV. 
Combined order TV \cite{Combined_order_TV} is another multi-order formulation proposed as a simple combination
of first and second order TV functionals through user-tuned coupling parameters and 
demonstrated performance comparable to the state of art image restoration 
techniques including TGV2. 
	
The success  of using multiple order derivatives lies with the fact that the  relative 
distribution  of derivative magnitude of different orders   can be exploited to retain 
signal sharpness in the  presence of noise; this is achieved by choosing appropriate
relative weights for derivative magnitudes of different orders.  However,  these
weights are applied only on the global level after pixel-wise summation of the
derivative magnitudes.  In 2D,  for a given pixel location,  and given order,  this 
magnitude is the norm of the result of vector differential operation. To elaborate on this
with an example,  consider the vectors derivative operators,   
${\bf L}_1 = [\frac{\partial}{\partial x}\;\;\frac{\partial}{\partial y} ]$, 
${\bf L}_2 = [\frac{\partial^2}{\partial x^2}\;\;\frac{\partial^2}{\partial x^2}\;\;
\sqrt{2}\frac{\partial^2}{\partial x\partial x} ]$.  
Then the combined order TV functional of Lysaker et al.  
\cite{Combined_order_TV} can be expressed  as
\begin{equation}
\label{eq:covlys}
R_{ctv}(g) = \alpha_1 \sum_{\bf r} \left\|({\bf L}_1g)({\bf r})\right\|_2 + 
 \alpha_2 \sum_{\bf r} \left\|({\bf L}_2g)({\bf r})\right\|_2 
\end{equation}
The method of Bredies et al \cite{Bredies_TGV_2010} uses a different
approach to combine orders \cite{Bredies_TGV_2010}, but the contributions from derivative orders are combined in the same way as above:  the weighting is applied
 after summing the pixel-wise derivative magnitudes.  Further, all 
such methods leave the relative weights ($\alpha_1$ and $\alpha_2$) as user parameters.

\subsection{Contribution and Outline}	 

We are interested in developing a multiple order derivative  based
regularization functional where re-combining is performed directly
on the derivative values without taking the magnitudes.  For simplicity,
we explore 1D signal restoration, where we consider the following  way of combining
multiple order derivatives:
\begin{equation}
\label{eq:moreg1}
R(g) = \sum_{r} \left\|{\bf S}(Lg)(r)\right\|_2,
\end{equation}
where
${\bf L} = \left[ \frac{\partial}{\partial x}, \; \frac{\partial^2}{\partial x^2},
 \; \cdots \; \frac{\partial^K}{\partial x^K}\right]^T$ is vector derivative operator,  and
${\bf S}$ is a $K\times K$ full rank matrix.  Our goal is to develop a probabilistic
framework  to construct a signal restoration method using the above form of
regularization with automatic determination of the matrix ${\bf S}$.  
Our contributions can be  summarized as given below:
\begin{enumerate}
\item
 We introduce a novel multivariate Laplacian density to model multiple order 
derivatives as well as their inter-dependencies. We then derive the regularization 
functional corresponding to the proposed density, where the  interdependency is 
modeled by a symmetric positive definite matrix, which we call as the structure matrix ${\bf S}$. 
\item
We derive the proposed regularization applied on a signal $g$, 
as the cross entropy measure between the proposed multivariate Laplacian model, and
sample multivariate density function constructed from the multi-order derivatives of the signal $g$.  We call this regularization
functional as the generalized multi-order total variation functional (GMO-TV). 
\item
Given an example signal $g$,  we derive a majorization-minimization 
algorithm to determine ${\bf S}$ as the minimizer of KL divergence between
the proposed multivariate Laplacian prior probability  density modeled by ${\bf S}$ and
sample multivariate density function constructed from the multi-order derivatives of the signal $g$.  Interestingly, the cost minimized
here is mostly identical to GMO-TV.  We call this algorithm the MM-KL.
\item
Next, we develop a training-based signal restoration method.
Suppose we are given a set of noise-free training signals, 
$\{g_1,\ldots,g_n\}$ and the noisy measured signal $f$ originating from an
 underlying signal. We need to estimate this signal, which  belongs to
 the class represented by the model signals $\{g_1,\ldots,g_n\}$. The
first  method  determines ${\bf S}$ from  $\{g_1,\ldots,g_n\}$ using MM-KL
and  restores the  required signal from $f$ by minimizing GMO-TV  functional
parameterized by ${\bf S}$.  The minimization for $f$ is also carried out
using majorization-minimization approach. We call this algorithm the
MM-GMOTV.
\item
 Further,  we develop a signal restoration method in which the required signal
 and the structure matrix are jointly determined as the minimizers of
 GMO-TV.  The proposed method is constructed as an alternation between
 MM-KL and MM-GMOTV.  We also provide a proof of convergence for this 
 method.   
  \item 
We apply the proposed approach for 1D signal restoration on ECG and EEG 
signals including denoising and deblurring. The experimental results show that the 
proposed methods perform better than current multi-order derivative based methods,
while having no requirement for tuning the model parameters as opposed to the compared
methods, which leave the model parameters as user parameters.
\end{enumerate}

The rest of the paper is as follows: Section 2 presents the signal restoration problem,
in two view points:  maximum a posteriori estimation and cross entropy penalized
maximum likelihood estimation.  Section 3 presents   the proposed multi-variate 
 Laplacian prior parameterized by a structure matrix and the derivation of GMO-TV
 functional.   It also develops an algorithm
 to determine the structure matrix (MM-KL).    Section 4 presents the training-based
 signal restoration using the GMO-TV  functional (MM-GMOTV).  Section 5 presents the
 training-free
 signal restoration approach using the GMO-TV  functional.
 Section 6  presents some experimental results.

\section{Signal restoration as MAP estimation and Cross-entropy penalized
ML estimation}
\label{sec:mapcemle}
	
Let $g_{o}:\Omega \to \mathbb{R}$ be the original uncorrupted 1D discrete signal
 defined on finite $\Omega \subseteq \mathbb{Z}$ and $f:\Omega \to \mathbb{R}$ 
 be the discrete measurement of $g_{o}$ given by
\begin{equation}
\label{eq:blurring_model}
f(x) = h(x) \ast g_{o}(x) + \eta(x), \ x \in \Omega
\end{equation}
where $h$ is a known linear transfer function representing the distortion and 
$\eta$ is the additive noise that corrupts the measurement. When $h$ is  
considered to be  $\delta(x)$,  the restoration problem becomes denoising.  It is given that the pdf of distribution of noise $\eta(x)$
is $p_Y(y,m)$,  where $m$ denotes the ideal measurable value and $y$ denotes
the actual value measured by the acquisition device.    It is also given that the values
of the derivatives of $g_o$ (specific order) follows a distribution with pdf  $q_V(v)$.
With these definitions, the conditional probability for $f$ being the measured signal
 with the condition that the given candidate signal $g$ is the source of the measurement,
 can be expressed as 
 \begin{equation}
 \label{eq:pfg}
 p(f|g) = \prod_x p_Y(f(x),(h*g)(x)),
 \end{equation}
 where we have assumed that, for any two index $x_1$ and $x_2$,  the noise
  is independently distributed.

The posterior probability of a candidate signal $g$, given measurement $f$ is 
\begin{equation}
\label{eq:map1}
p(g|f)=\frac{p(f|g)p(g)}{p(f)},
\end{equation}
where $p(f|g)$ is the probability of obtaining the observation $f$  given $g$, $p(g)$  
is the probability distribution of $g$ and $p(f)$ is the probability distribution of $f$. 
The maximum a posteriori method (MAP)  computes the required signal as a maximizer 
of the above probability with respect to $g$ as the  maximization variable.  Since  $p(f)$ is 
independent of the maximization variable $g$, it can be skipped from the expression. 
The probability  $p(g)$ is known as prior probability and is defined in terms of point-wise 
roughness of the signal. To maximize the above probability,  we minimize its negative 
logarithm. The MAP based signal restoration amounts to finding $\hat{g}$ as a minimizer
of negative logarithm of $p(g|f)$.  In the remainder of the paper,  we restrict ourself
to Gaussian pdf for $p_Y(\cdot,m)$ which is the most commonly used assumption.

Our focus is now on investigating the form of 	 $p(g)$ such that negative log of $p(g|f)$
becomes compatible with known forms of cost functions used in the literature for
signal restoration.  $p(g)$ can be expressed   as 
\begin{equation}
\label{eq:pg}
p(g) = \prod_x q_V((L*g)(x)),
\end{equation}
where $L$ is the discrete filter implementing the derivative of a 
given order, and 
$q_V(\cdot)$ is the prior probability on the distribution of derivatives.
 Here too,
we assume that the distributions of derivatives across different sample locations
are independent.  With this assumption,  the negative log of $p(g|f)$ can be written
as follows with $p_Y(\cdot,m)$ restricted to be Gaussian pdf:
\begin{equation}
\label{eq:jgen}
J(g) = \underbrace{\frac{1}{2}\sum_x (((h*g)(x) - f(x))^2}_{D(f,g)}
  \underbrace {-\sum_x  log(q_V((L*g)(x)))}_{R(g)}.
\end{equation}
 The independence assumption on the derivatives at different sample locations is  
 clearly not true.  However,  well-known cost functionals used for signal restoration can 
 be expressed using this assumption.  For example,  the following form of 
 $q_V(\cdot)$ will give the well-known quadratic functional,
\begin{align}
	q_{\V}(\v) = \frac{1}{Z}e^{\frac{-\lambda||\v||_{2}^{2}}{2}}, \v \in {\mathbb{R}}.
\end{align}
This quadratic functional is known as Tikhonov functional, which
can be expressed as
\begin{equation}
\label{eq:tikh}
R(g) =\lambda \sum_x  ((L*g)(x))^2.
\end{equation}
Next, the total variation functional 
\begin{equation}
\label{eq:tv}
R(g) =\lambda \sum_x  |((L*g)(x))|,
\end{equation}
 is the result of using the following form of $q_V(\cdot)$:
\begin{align}
	q_{V}(v) = \frac{1}{Z}e^{\frac{-\lambda||\v||_{2}}{2}}.
\end{align}

An alternative way 	to get the form in the equation (\ref{eq:jgen}) is by summing
negative log of $p(g|f)$ with a penalty term known as the cross entropy measure.  Specifically,
\begin{equation}
\label{eq:jgen2}
J(g) = \frac{1}{2}{\sum_x (((h*g)(x) - f(x))^2}  + 
H(p_{\scriptscriptstyle V, g},q_{\scriptscriptstyle V}),
\end{equation}
where   $H(p_{\scriptscriptstyle V, g},q_{\scriptscriptstyle V})$ is the cross entropy 
measure between $q_{\scriptscriptstyle V}$ and the sample pdf  obtained from the
derivatives of the candidate signal $g$.  The cross entropy can be expressed as given
below:
\begin{equation}
\label{eq:hpq}
H(p_{\scriptscriptstyle V, g},q_{\scriptscriptstyle V}) =
-\int_{\bf v} p_{\scriptscriptstyle V, g}({\bf v})
log [q_{\scriptscriptstyle V}({\bf v})]d{\bf v}
\end{equation}
The sample pdf $p_{\scriptscriptstyle V, g}$ is expressed in the form of a 
Parzen window based estimator as given below: 
\begin{equation}
\label{eq:pvgparzen}
p_{\scriptscriptstyle V, g}(\mathbf{v)} =
 \frac{1}{Z}\sum_{x} G_{\sigma}(||\mathbf{v}-(L*g)(x)||_{2}^{2}) \\
\end{equation}
where $G_{\sigma}$ is the Gaussian kernel with size $\sigma$, and $Z$ is a
normalization constant.  Substituting equation (\ref{eq:pvgparzen}) in the
equation (\ref{eq:hpq}) gives
\begin{equation}
\label{eq:hpq2}
H(p_{\scriptscriptstyle V, g},q_{\scriptscriptstyle V}) =
-\int_{\bf v}  \frac{1}{Z}\sum_{x} G_{\sigma}(||\mathbf{v}-(L*g)(x)||_{2}^{2}) 
log [q_{\scriptscriptstyle V}({\bf v})]d{\bf v}
\end{equation}
Taking the integral inside the summation gives
\begin{equation}
\label{eq:hpq3}
H(p_{\scriptscriptstyle V, g},q_{\scriptscriptstyle V}) =
-  \frac{1}{Z}\sum_{x} \int_{\bf v} G_{\sigma}(||\mathbf{v}-(L*g)(x)||_{2}^{2}) 
log [q_{\scriptscriptstyle V}({\bf v})]d{\bf v}.
\end{equation}
To simplify further,  we take the limit $\sigma \rightarrow 0$.  With this limit,
the integral $\int_{\bf v} G_{\sigma}(||\mathbf{v}-(L*g)(x)||_{2}^{2}) 
log [q_{\scriptscriptstyle V}({\bf v})]d{\bf v}$ becomes 
$log [q_{\scriptscriptstyle V}((L*g)(x)]$  because
$G_{\sigma}(||\mathbf{v}-(L*g)(x)||_{2}^{2})$ becomes a sampling kernel for
$\sigma \rightarrow 0$. 
  Hence,   the cross entropy becomes
\begin{equation}
\label{eq:hpq4}
H(p_{\scriptscriptstyle V, g},q_{\scriptscriptstyle V}) =
-  \sum_{x} 
log [q_{\scriptscriptstyle V}((L*g)(x))].
\end{equation}
This means that the $H(p_{\scriptscriptstyle V, g},q_{\scriptscriptstyle V})$
is identical to  $R(g)$  introduced in the equation (\ref{eq:jgen}).  Hence,
cross entropy augmented negative log of   data-Likelihood  is identical
to negative log of posterior probability,  and hence cross entropy
penalized ML estimation is the same as the MAP estimation.
Needless to say, the well-known Tikhonov and total variation functionals of
equations (\ref{eq:tikh}), and (\ref{eq:tv}) can also be derived as specific cases of the cross entropy $H(p_{\scriptscriptstyle V, g},q_{\scriptscriptstyle V})$.

So far, we have reviewed two formulations that lead to cost functionals used for regularized
signal restoration namely, MAP estimation,  and cross entropy penalized ML (CE-ML) estimation.
The advantage of the latter formulation is that,  it does not assume that signal derivatives
across difference sample locations are distributed independently.  In the following section, we propose the 
GMO-TV functional parameterized by the so-called structure matrix,  and derive  an iterative algorithm
 to determine the  structure matrix using cross entropy formulation.

\section{Generalized Multi-Order Total Variation Functional}
	
\subsection{Multivariate Laplacian prior on signal derivatives
and the corresponding regularization functional}

Consider a signal $g(x)$  with derivative $\mathbf{v}_{g}(x)$ given
by $\mathbf{v}_{g}(x) = \mathbf{L}(x) \ast g(x)$,
where ${\bf L}$ is a vector derivative filter for derivatives upto the 
$K^{th}$ order and given by 
\begin{equation}
\label{eq:vector_derivative_operator}
{\bf L}(x) = \left[ \frac{\partial}{\partial x}, \; \frac{\partial^2}{\partial x^2}, 
\; \cdots \; \frac{\partial^K}{\partial x^K}\right]^T,
\end{equation}
To incorporate a general prior best suited for modeling long-tailed distribution
observed in signal derivatives and also handle local inter-dependencies
 among these derivatives, we propose to use the following form of multivariate
  Laplacian prior:
\begin{equation}
\label{eq:laplacian_prior_pdf}
q_{\mathbf{V}}(\mathbf{v}) =  \frac{1}{Z(K) |{\bf C}|^{\frac{1}{2}}}
e^{-\sqrt{\left( {\bf v}^T{\bf C}^{-1} {\bf v} \right)}}, 
\mathbf{v} \in \mathbb{R}^{K},
\end{equation} 
where ${\bf C}$ is positive definite matrix, and $Z(K)$ is normalization constant.
Next, the sample multivariate 	pdf estimated from the multi-order 
derivatives of $g$ can be expressed as
\begin{equation}
\label{eq:pvgparzen2}
p_{\scriptscriptstyle V, g}(\mathbf{v)} =
 \frac{1}{Z_p}\sum_{x} G_{\sigma}(||\mathbf{v}-(L*g)(x)||_{2}^{2}), \\
\end{equation}		
where $Z_p$ is another normalization constant.
Now,   note that the steps used in the second part of Section \ref{sec:mapcemle}
to derive the expression for $H(p_{\scriptscriptstyle V, g},q_{\scriptscriptstyle V})$
in the univariate case are directly extendible for the multivariate case.   Hence
$H(p_{\scriptscriptstyle V, g},q_{\scriptscriptstyle V})$ for the current multivariate
case can be expressed as
\begin{equation}
\label{eq:hpqv}
H(p_{\scriptscriptstyle V, g},q_{\scriptscriptstyle V}) =
-  \sum_{x} 
log [q_{\scriptscriptstyle V}(({\bf L}*g)(x))].
\end{equation}
Substituting the expression of the equation (\ref{eq:laplacian_prior_pdf}) in the above
equation gives
\begin{equation}
\label{eq:hpqvs}
\bar{R}(g, {\bf C}) = H(p_{\scriptscriptstyle V, g},q_{\scriptscriptstyle V}) =
-  \sum_{x} 
\sqrt{(({\bf L}*g)(x))^T{\bf C}^{-1}({\bf L}*g)(x)}  + \frac{1}{2} \log |{\bf C}|.
\end{equation}
 Here we have ignored the constant that is independent of both
 $g$ and ${\bf C}$.  As a regularization functional applied on $g$,  we ignore the term involving only ${\bf C}$
and write
\begin{equation}
\label{eq:gmotv}
R(g,{\bf C}) =
 \sum_{x} 
\sqrt{(({\bf L}*g)(x))^T{\bf C}^{-1}({\bf L}*g)(x)}.
\end{equation}
Since ${\bf C}^{-1}$ is a symmetric matrix, it can be written as 
${\bf C}^{-1} = {\bf U}{\bf D}{\bf U}^T$, where ${\bf U}$ is an orthonormal matrix
satisfying ${\bf U}^T{\bf U} = {\bf I}$, and ${\bf D}$ is a diagonal matrix.  Hence
${\bf C}^{-1}$ can be written as ${\bf C}^{-1} ={\bf S}^T{\bf S}$, where 
${\bf S}$ is a matrix of the form 
$${\bf S} = \left[\begin{array}{c}  {\bf p}_1^T \\ {\bf p}_2^T \\ \ldots,\\ {\bf p}_k^T \end{array} \right],$$
wtih ${\bf p}_i$'s satisfying  ${\bf p}_i^T{\bf p}_j=0$ for $i\ne j$, and
 ${\bf p}_i^T{\bf p}_i> 0$.  Substituting for $\C^{-1}$ in terms of $\S$, we get the corresponding functional $R(g,{\bf S})$ as
 \begin{equation}
\label{eq:gmotv2}
R(g,{\bf S}) =
 \sum_{x} 
\|{\bf S} ({\bf L}*g)(x)\|_2.
\end{equation}
Since $|{\bf C}|=\frac{1}{|{\bf S}^T{\bf S}|}=\frac{1}{|{\bf S}{\bf S}^T|}$, 
$\bar{R}(g, {\bf S})$ can be written as
\begin{equation}
\label{eq:hpqv2}
\bar{R}(g, {\bf S}) =  \sum_{x} 
R(g,{\bf S})  - \frac{1}{2} \log |{\bf S}{\bf S}^T|.
\end{equation}

\subsection{Determining the structure matrix ${\bf S}$}

Minimizing derivative based roughness functional helps to suppress noise;  but it also leads
to the loss of resolution since sharp signal variations are suppressed by minimizing 
derivative
magnitude.  In this viewpoint,  the purpose of introducing multi-order derivative is the following:
instead of minimizing individual derivative magnitudes,   we intend to minimize the deviation
from the pre-determined inter-relationship among derivatives of multiple order, which will
hopefully reduce the loss of signal resolution.  The inter-relationship is captured by 
$q_{\mathbf{V}}(\mathbf{v})$ by means of the structure matrix,  and its deviation from the
inter-relationship present in the candidate signal $g$ is measured by the cross entropy,
 $H(p_{\scriptscriptstyle V, g},q_{\scriptscriptstyle V})$.  

Here, we address the problem of determining ${\bf S}$, given an example signal $\bar{g}$.
An obvious approach is to determine ${\bf S}$ by minimizing 
$H(p_{\scriptscriptstyle V, g},q_{\scriptscriptstyle V})$  given in equation (\ref{eq:hpqv}).
This approach is also optimal  in information theoretic viewpoint:  it minimizes the
complexity of representing the derivatives of $g$  using the pdf parameterized by ${\bf S}$
by means of cross entropy measure.  Interestingly,  the well-known KL divergence  which
is also used to estimate a parametric pdf from given set of samples coincide with 
cross entropy measure.  The KL divergence between the pdf's 
$p_{\scriptscriptstyle V, g}$ and $q_{\scriptscriptstyle V}$  is expressed as
\begin{align}
D_{KL}(P||Q) = &\int_{\bf v} p_{\scriptscriptstyle V, g}({\bf v}) 
                         \log\left(\frac{p_{\scriptscriptstyle V, g}({\bf v}) }{q_{\scriptscriptstyle V}(\bf v)}\right) d{\bf v}\\
	             = &\int_{\bf v} p_{\scriptscriptstyle V, g}({\bf v})  \log(p_{\scriptscriptstyle V, g}({\bf v}) ) d{\bf v}
	             - \int_{\bf v}p_{\scriptscriptstyle V, g}({\bf v}) \log(q_{\scriptscriptstyle V}(\bf v)) d{\bf v}\\
	             = & \ -H(p_{\scriptscriptstyle V, g}) + H(p_{\scriptscriptstyle V, g},q_{\scriptscriptstyle V})
\end{align}
Since $H(p_{\scriptscriptstyle V, g})$ is independent of ${\bf S}$, this means that minimizing
$D_{KL}(P||Q)$  with respect to ${\bf S}$ is the same as  
minimizing $H(p_{\scriptscriptstyle V, g},q_{\scriptscriptstyle V})$.   In the following section, 
we develop a majorization-minimization method for determining ${\bf S}$ by minimizing
$D_{KL}(P||Q)$. 

Now our goal is to develop a computational method for the following minimization problem:
\begin{equation}
\label{eq:probkl}
{\bf S}^* = \arg \min_{\bf S } \bar{R}(g, {\bf S}) = \arg \min_{\bf S }
{R}(g, {\bf S})    - \frac{1}{2} \log |{\bf S}{\bf S}^T|.
\end{equation}
where $R(g,{\bf S})$ is as defined in equation (\ref{eq:gmotv2}).  In order to make
this method useful for both training-based and training-free signal restoration methods
(Sections 3 and 4),  we need to modify the above problem as given below:
\begin{equation}
\label{eq:probklf}
{\bf S}^* = \arg \min_{\bf S } {R}_F(g, {\bf S}) = \arg \min_{\bf S }
{R}(g, {\bf S})    - \frac{1}{2} \log |{\bf S}{\bf S}^T| + \lambda_F \|{\bf S}\|_F,
\end{equation}
where  $\|\cdot\|_F$  denotes the Frobenius norm of its matrix argument.
The reason for using ${R}_F(g, {\bf S})$ instead of  $\bar{R}(g, {\bf S})$
is that ${R}_F(g, {\bf S})$ is bounded below for all $g$, while $\bar{R}(g, {\bf S})$
is not bounded below if the derivatives of $g$ are not well-distributed.

Let
${\bf p}^T = [{\bf p}_1^T\cdots {\bf p}_k^T]$. Note that ${\bf p}_i^T$'s are the rows of
${\bf S}$.  Let ${\bf v}(x) =  ({\bf L}*g)(x)$.  Let ${\bf S}_0$ be the initialization towards
iteratively solving the above problem.  To solve the above computational problem,
we adopt majorization-minimization approach.
Given current estimate of the minimum, say ${\bf S}^{(k)}$,  we build an ${\bf S}^{(k)}$-dependent
auxiliary functional,  ${R}_F^{(k)}(g, {\bf S}, {\bf S}^{(k)})$ satisfying 
 ${R}^{(k)}_F(g, {\bf S}^{(k)}, {\bf S}^{(k)})$ $={R}_F(g, {\bf S}^{(k)})$,  and
 ${R}^{(k)}_F(g, {\bf S}, {\bf S}^{(k)}) > {R}_F(g, {\bf S}^{(k)})$ for 
 ${\bf S} \ne {\bf S}^{(k)}$.  Then we compute the
 next refined estimate as
 \begin{equation}
 \label{eq:mmjgss}
 {\bf S}^{(k+1)} = \arg\min_{\bf S} {R}^{(k)}_F(g, {\bf S}, {\bf S}^{(k)}).
 \end{equation}
 To construct the  majorizer for  ${R}_F(g, {\bf S})$,
 we need to find the majorizer for ${R}(g, {\bf S})$ which is the most
 complex part of ${R}_F(g, {\bf S})$.  The ${\bf S}^{(k)}$-dependent
 majorizer for  ${R}(g, {\bf S})$, denoted by ${R}^{(k)}(g, {\bf S}, {\bf S}^{(k)})$
 can be expressed as
 \begin{equation}
 \label{eq:rgsmaj}
 {R}^{(k)}(g, {\bf S}, {\bf S}^{(k)}) = \sum_x   
 \frac{0.5}{\|{\bf S}^{(k)}({\bf L}*g)(x)\|_2}\|{\bf S}({\bf L}*g)(x)\|_2^2.
 \end{equation}
 Based on this, the majorizer for ${R}_F(g, {\bf S})$  can be written as
  \begin{equation}
 \label{eq:rgsbmaj}
 {R}^{(k)}_F(g, {\bf S}, {\bf S}^{(k)}) = \sum_x   
 \frac{0.5}{\|{\bf S}^{(k)}({\bf L}*g)(x)\|_2}\|{\bf S}({\bf L}*g)(x)\|_2^2 - \log|{\bf S}^T{\bf S}|
 + \lambda_F \|{\bf S}\|_F
 \end{equation}
 
 To construct the algorithm based on the above majorization,  we will need
 the expression for the gradients of ${R}^{(k)}_F(g, {\bf S}, {\bf S}^{(k)})$
 and ${R}_F(g, {\bf S})$, which are given in the following proposition.
 For notational convenience,
we represent this gradient in matrix form with same size as ${\bf S}$.
\begin{prop}
The gradient  $\nabla_{\bf S}{R}_F(g,{\bf S})$  is given by
\begin{equation}
\label{eq:gradjs}
\nabla_{\bf S}{R}_F(g,{\bf S})  ={\bf S}{\bf A} - 
 ({\bf S}{\bf S}^T)^{-1}{\bf S} + \lambda_F {\bf S}
\end{equation}
where ${\bf A} = \sum_{x} \frac{1}{\|{\bf S}({\bf L}*g)(x))\|_2}(({\bf L}*g)(x))(({\bf L}*g)(x))^T$ \\
The gradient $\nabla_{\bf S}{R}^{(k)}_F(g,{\bf S},{\bf S}^{(k)})$  is given by
\begin{equation}
\label{eq:gradjsm}
\nabla_{\bf S}{R}^{(k)}_F(g,{\bf S},{\bf S}^{(k)})  =  {\bf S}{\bf A}_k - 
 ({\bf S}{\bf S}^T)^{-1}{\bf S} +  \lambda_F {\bf S},
\end{equation}
where ${\bf A}_{k} = \sum_{x}\frac{1}{\|{\bf S}^{(k)}(({\bf L}*g)(x))\|_2}(({\bf L}*g)(x))(({\bf L}*g)(x))^T$
\end{prop}
Based on the gradient expression for ${R}^{(k)}_F(g, {\bf S}, {\bf S}^{(k)})$ from the
above proposition,  we get the closed form expression for the minimum of 
 ${R}^{(k)}_F(g, {\bf S}, {\bf S}^{(k)})$  with respect to ${\bf S}$, which is given in the following
 proposition.
 \begin{prop}
 The minimum of  ${R}^{(k)}_F(g, {\bf S}, {\bf S}^{(k)})$  with respect to ${\bf S}$ 
 is given by 
 ${\bf S}^{(k+1)}= ({\bf D}_k+\lambda_F{\bf I})^{-1/2}
  {\bf U}_k^T$, where  ${\bf D}_k$  and ${\bf U}_k$
 are matrices involved in the Eigen decomposition of ${\bf A}_{k}$, i.e., 
  ${\bf A}_{k}  = {\bf U}_k{\bf D}_k {\bf U}_k^T$.
 \end{prop}
 Together with MM scheme expressed by equation (\ref{eq:mmjgss}), 
 this completes the derivation of iterative algorithm for minimizing ${R}_F(g, {\bf S})$.
For readers' convenience,  we  express the full  algorithm in terms of computational steps.
The input ${\bf v}$  is given by  ${\bf v}(x) = ({\bf L}*g)(x)$ \\
{\em Algorithm I: \;\; $\textrm{MM-KL}({\bf v}, {\bf S}^{(0)}, \lambda_F, \epsilon)$}
\begin{align*}
Initialization:   {\bf A}_{0}   &
 = \sum_{x} \frac{1}{\left\|{\bf S}^{(0)}{\bf v}(x)\right\|_2}{\bf v}(x){\bf v}^T(x),
  \; k = 0,  \; r = 1\;\;\;\;\;\;\;\;   \\
while \;\;\;\;\;   r   > \epsilon \;\; do  &  \\
& Factorize \;\; {\bf A}_{k}  = {\bf U}_k{\bf D}_k {\bf U}_k^T \\
& Update \;\;  {\bf S}^{(k+1)}  = ({\bf D}_k+\lambda_F{\bf I})^{-1/2} {\bf U}_k^T,  \\
& Compute\;\; {\bf A}_{k+1}   = 
\sum_{x} \frac{1}{\left\| {\bf S}^{(k+1)}{\bf v}(x)\right\|_2}{\bf v}(x){\bf v}^T(x) \\
 &  k \leftarrow k+1 \\
 & r  =  \|{\bf S}^{(k)}{\bf A}_k - 
 ({\bf S}^{(k)}({\bf S}^{(k)})^T)^{-1}{\bf S}^{(k)} + \lambda_F{\bf S}^{(k)} \|_2\\
 \;\; Else\;\; return \;\; {\bf S}^{(k)}.
\end{align*}
Clearly, MM-KL converges to the minimum of  ${R}_F(g, {\bf S})$
 for $\epsilon = 0$. However, for practical purposes, we use small positive value
 $\epsilon$. Note that, calling MM-KL with $\lambda_F=0$ returns the
 minimum of $\bar{R}(g, {\bf S})$, and returns the minimum of ${R}_F(g, {\bf S})$
 otherwise. Note that $R(g,\S)$ and $R_{F}(g,\S)$ are both non-differentiable when $\v(x)$ = $(\mathbf{L}\ast g)(x)$ = $\mathbf{0}$. For handling such cases, we replace  $\|{\bf S}^{(k)}({\bf L}*g)(x)\|_2$ with the approximation
 $\sqrt{\epsilon + ||{\bf S}^{(k)}({\bf L}*g)(x)||_{2}^{2}}$ in practice, with $\epsilon$ as a small positive constant. 
For notational convenience, we use the term $\|{\bf S}^{(k)}({\bf L}*g)(x)\|_2$ in equations, while differentiability is retained through the above mentioned approximation.

\section{GMO-TV based signal restoration with training}

  Suppose we have a set of noise-free training signal models, 
$\{g_1,\ldots,g_n\}$ and the noisy measured signal $f$ originating from an
 underlying signal, which we need to estimate, and which  belongs to
 the class represented by the model signals $\{g_1,\ldots,g_n\}$.  
 Let ${\bf v}(x)$ denote the vector sequence obtained by augmenting
 the vector sequences $\{({\bf L}*g_j)(x)\}_{j=1,\ldots,n}$ across the
 index $x$.   We first determine  ${\bf S}$ from ${\bf v}(x)$, 
 by calling   MM-KL with ${\bf S}^{(0)} = {\bf I}$,
 $\lambda_F = 0$, and with a sufficiently low value for $\epsilon$.
 Let ${\bf S}^*$  be the result returned by MM-KL.
 We then get the restored signal from $f$ by minimizing the following
 cost:
 \begin{equation}
 \label{eq:jgsp}
 \bar{J}(g,{\bf S}^*) = \frac{1}{2}\sum_{x} (f(x) - (h \ast g)(x) )^2 + \lambda \bar{R}(g,{\bf S}^*).
 \end{equation}
 Note that, with respect to $g$,  the cross entropy, $\bar{R}(g,{\bf S}^*)$, differs 
 from ${R}(g,{\bf S}^*)$ only by a constant. 
 Hence, we can as well minimize the following cost to get the required signal:
 \begin{equation}
 \label{eq:jgsp2}
 J(g,{\bf S}^*) = \frac{1}{2}\sum_{x} (f(x) - (h \ast g)(x) )^2 + \lambda {R}(g,{\bf S}^*).
 \end{equation}
 
 To express the gradient,  we first define the following weight sequence
 based on a given signal $\bar{g}$:
 \begin{equation}
 \label{eq:wg}
 w_{ [\bar{g},{\bf S}^*]}(x) = \frac{1}{2\left({ \| {\bf S}^*  \left(\mathbf{L}(x) 
 \ast \bar{g}({x}) \right)\|_{2}}\right)}.
\end{equation}
Based on this,  we define the following $\bar{g}$-dependent operator on signal
$g$:
\begin{equation}
\label{eq:Qg}
({\cal Q}_{ [\bar{g},{\bf S}^*]}g)(x) =  h(-x) \ast h(x)\ast g(x)  + 
 \lambda \mathbf{L}^{T}(x) \ast \left\{ w_{[\bar{g},{\bf S}^*]}(x)\mathbf{S}^{*T}
\mathbf{S}^*\left[ \mathbf{L}g(x)\right] \right\}
\end{equation}
Note that ${\cal Q}_{ [\bar{g},{\bf S}^*]}\left[g(x)\right]$ is a linear operator on $g$
if $g\ne \bar{g}$.  If $\bar{g}$ is replaced with $g$, it becomes a non-linear
operator, i.e.,  ${\cal Q}_{ [g,{\bf S}^*]}\left[g(x)\right]$ is a non-linear operator on
$g$.  Now the gradient of $J$ at a given candidate signal $g$ can be expressed
as
\begin{equation}
\label{eq:gradjgs}
d_{[g,{\bf S}^*]}(x) = \nabla_g J(g,{\bf S}^*) = ({\cal Q}_{[g,{\bf S}^*]}g)(x)  - h(-x) \ast f(x),
\end{equation}
where the subscript in $\nabla_g$ signifies that fact that gradient is taken with
respect to $g$.  Note that $\nabla_g J(g,{\bf S}^*)$ is the collection of derivatives
with respect to each sample or element of $g$, and its number of elements
is the same as that of $g$.  Hence we represent the gradient using notational
form used for signal, i.e.,  we denote the gradient by $d_{[g,{\bf S}^*]}(x)$.

The minimum for $J(g,{\bf S}^*)$ can be obtained by solving $\nabla_g J(g,{\bf S}^*) = 0$.
We can either use majorization-minimization
(MM) approach,  or nested nonlinear conjugate gradient approach  (NNCG) \cite{NNCG_Deepak} for
minimizing $J(g,{\bf S}^*)$.   Although NNCG is faster than  MM approach,  the difference 
in speed will be insignificant since the current problem is in 1D; on the other hand,  MM method
is easier to implement.  Hence use the MM approach here. 

Given current estimate of minimum, say $g^{(k)}$,  we build a $g^{(k)}$-dependent
auxiliary functional,  $J^{(k)}(g, {\bf S}^*, g^{(k)})$ satisfying $J^{(k)}(g^{(k)}, {\bf S}^*, g^{(k)})
= J(g^{(k)}, {\bf S}^*)$  and 
 $J^{(k)}(g, {\bf S}^*, g^{(k)}) >  J(g^{(k)}, {\bf S}^*)$ for $g \ne  g^{(k)}$.   Then we compute the
 next refined estimate as
 \begin{equation}
 \label{eq:mmjgsg}
 g^{(k+1)} = \arg\min_{g} J^{(k)}(g, {\bf S}^*, g^{(k)}).
 \end{equation}
 $J^{(k)}(g, {\bf S}^*, g^{(k)})$ is constructed as given below:
 \begin{equation}
\label{ICcostq}
J^{(k)}(g, {\bf S}^*,g^{(k)})  =  \sum_{x} (f(x) - (h \ast g)(x) )^2  + 
 \lambda \sum_{x} w_{ [g^{(k)},\S^*]}(x) { || {\bf S}^*  \left(\mathbf{L}(x) \ast g({x}) \right) ||^{2}}
\end{equation}
 The gradient of above cost is given by
 \begin{equation}
 \label{eq:gradjgsq}
d^{(k)}_{[g,{\bf S}^*]}(x)  =
 \nabla_g J^{(k)}(g, {\bf S}^*,g^{(k)}) = ({\cal Q}_{[g^{(k)},\mathbf{S}^*]}g)(x)  - h(-x) \ast f(x).
 \end{equation}
 The minimum of $J^{(k)}(g, {\bf S}^*, g^{(k)})$  can be computed by solving 
 $\nabla_g J^{(k)}(g, {\bf S}^*,g^{(k)})$ = $0$.

 The cost $J^{(k)}(g, {\bf S}^*,g^{(k)})$   itself has to be solved iteratively,  i.e.,  each step
 in the MM update from $g^{(k)}$  to  $g^{(k+1)}$ defined in equation 
 (\ref{eq:mmjgsg}) should be solved iteratively.
 This  is   solved using the method of
 conjugate gradient (CG).  Let  $\{g^{(k)}_l\}_{l=0,1,\ldots}$ denote the sequence
 of iterates generated by this iteration.  Let $g^{(k)}$ be the initialization
 for this CG iteration, i.e.,  $g^{(k)}_0 = g^{(k)}$.  If $l^*$ denotes the index
 at which the termination criterion is attained,  $g^{(k+1)}$ becomes
 $g^{(k+1)}=g^{(k)}_{l^*}$.  We propose to use the following termination
 condition:  $\|\nabla_g J^{(k)}(g^{(k)}_{l^*}, {\bf S}^*,g^{(k)})\|_2 < \epsilon_q$,
 where  $\epsilon_q$ is a user-specified real positive number.
 Further,  we propose to terminate the MM iteration specified by equation
 (\ref{eq:mmjgsg})   on the attainment of condition    
 $\|\nabla_g J(g^{(k)}, {\bf S}^*)\|_2 < \epsilon_m$, where $\nabla_g J(g^{(k)}, {\bf S}^*)$ is as
 given in equation (\ref{eq:gradjgs}), and $\epsilon_{m}$ is an another
 user-specified real positive number. If the MM loop of equation
 (\ref{eq:mmjgsg}) is initialized with $g^{(0)}$,  and $\hat{g}$ is the result returned
 by the overall MM method with termination tolerances $\epsilon_q$ and $\epsilon_{m}$,
 we denote the action of the overall methods as 
 $\hat{g} = \textrm{MM-GMOTV}(g^{(0)}, {\bf S}^*,  \epsilon_q, \epsilon_{m})$. 
 
  To speed-up CG iterations, we use  the preconditioner based on the diagonal  
  approximation  of ${\cal Q}_{[g^{(k)},\S^*]}$.   This diagonal approximation
  is multiplication by the following: 
\begin{align}
{D}_{[g^{(k)}]}(x) = \ &\sum_{y} \left(h(y)\right)^{2} + 
\sum_{i=1}^{4} \left(\hat{\mathbf{L}}_{i}
(-x)\right)^{\boldsymbol{\cdot} 2} \ast w^{(k-1)}(x),
\end{align}
where $\left(\bullet\right)^{\boldsymbol{\cdot} 2}$  denotes the 
element-wise squaring of its filter argument,  and 
$\hat{\mathbf{L}}_{i}(x) = \ \mathbf{p}^T_{i}\mathbf{L}(x)$ with   
$\mathbf{p}^T_{i}$  being the $i^{th}$ row of $\mathbf{S}^*$.  The proposed
preconditioner is the division with ${D}_{[g^{(k)}]}(x)$.  As in previous section, we use the differentiable approximation for $|| {\bf S}^*  \left(\mathbf{L}(x) \ast g({x}) \right) ||_{2}$ in all cases.

Although the need for noise-free signals narrows-down the applicability of 
this method,  such scenarios are not unnatural.  For example,  in applications where
 signals such as ECG, EEG, are transmitted over a communication channel,
 one can compute ${\bf S}^*$  before transmission,  and use it to restore the signals
received through the transmission channel.  Another possibility is that,
one can compute ${\bf S}^*$ from signals acquired using expensive low-noise
equipments, and use it to restore signals that are acquired using inexpensive
noisy equipments.

 \section{GMO-TV based signal restoration without training}

\subsection{Eliminating the training}

To apply GMO-TV functional without the need for training signals,
we propose to formulate the signal restoration problem as a joint
minimization problem where both the signal, $g$, and the structure
matrix, ${\bf S}$, become minimization variables.
  Specifically,  the signal restoration becomes
as given below
  \begin{equation}
 \label{eq:jgspf}
(g^*,{\bf S}^*) = \arg\min_{(g,{\bf S})} 
 {J}_F(g,{\bf S}) = 
 \arg\min_{(g,{\bf S})} \frac{1}{2}\sum_{x} (f(x) - (h \ast g)(x) )^2 + 
 \lambda {R}_F(g,{\bf S}),
 \end{equation}
where ${R}_F(g,{\bf S})$ is as given in equation (\ref{eq:probklf}).
The restoration problem hence becomes estimating $g^*$ and ${\bf S}^*$
jointly such that  they agree with each other in the sense of cross
entropy,  and  $g^*$ fits the measured signal $f$ well.
Note that  ${R}_F(g,{\bf S})=\bar{R}(g,{\bf S})+\lambda_F\|{\bf S}\|_F$
 is essentially  the cross entropy 
$H(p_{\scriptscriptstyle V, g},q_{\scriptscriptstyle V})$ (except 
for the added Frobenius norm of ${\bf S}$),
where $q_{\scriptscriptstyle V}$ is the parametric pdf
expressed in terms of ${\bf S}$ (equation (\ref{eq:laplacian_prior_pdf})),
and $p_{\scriptscriptstyle V, g}$ is the sample pdf of the derivatives
of $g$ (equation (\ref{eq:pvgparzen})).  Note that  $\bar{R}(g,{\bf S})$
can become unbounded below  with respect to ${\bf S}$ when
$g$ does not have its derivatives sufficiently distributed (example: $g=0$).
This is why $\|S\|_F$ has been included, which makes the overall
cost bounded below even for the cases when $g=0$.  Note that
$\lambda_F$ can be chosen to be arbitrary low, and the boundedness
can still be ensured. 

To compute the solution for the above problem,  we adopt the method of
block coordinate descent.  Let ${\bf S}^{(0)}$ be the initialization.
Then the block coordinate descent method involves the following
steps  with $m$  being the iteration index:
\begin{align}
\label{eq:bcgmin}
\mbox{Step 1}:\;\;\;\;\;\; & g^{(m+1)}  =  \arg\min_{g} 
 {J}_F(g,{\bf S}^{(m)})\\
 \label{eq:bcsmin}
 \mbox{Step 2}:\;\;\;\;\; & {\bf S}^{(m+1)} =  \arg\min_{\bf S} 
 {J}_F(g^{(m+1)},{\bf S})
\end{align}
The algorithm expressed by equations (\ref{eq:bcgmin}) and (\ref{eq:bcsmin}),
belongs to the class of block-coordinate descent methods.  It is known that these
methods converge to a local minimum if the function is convex with respect to 
each block of variables  according to the result of Bertsekas \cite{Bertsekas_NLP}.   For our problem,  this
requirement is clearly satisfied, i.e.,  ${J}_F(g,{\bf S})$ is  convex either  with
respect to $g$ with ${\bf S}$ fixed, or with respect  to ${\bf S}$ with  $g$ fixed.
However,  these minimization sub-problems cannot be computed exactly as there are
no closed form solutions.  Hence the convergence results of Bertsekas is not
strictly applicable.    In the following section,  we first discuss about the iterative methods
for solving these sub-problems. Then, we propose  practical termination conditions
for the above minimization sub-problems 
to ensure convergence of the overall algorithm.

\subsection{Solving the subproblems}

 Note that, with respect to
$g$ alone, the functionals $J_F(g,{\bf S})$ and ${J}(g,{\bf S})$
differ only by a constant.  Hence, their gradients with respect to
$g$ are identical, i.e., $\nabla_g {J}_F(g,{\bf S}) = \nabla_g {J}(g,{\bf S})$,
which is given in equation (\ref{eq:gradjgs}).  Hence,  the minimization
in Step 1 can be solved by using    Majorization-Minimization method.
This can be done by calling MM-GMOTV  with $g^{(m)}$
as the initialization for the minimization variable $g$, and with ${\bf S}^{(m)}$
as parameter for GMO-TV functional.  In other words,  
 the result of step 1,  $g^{(m+1)}$, can be obtained as
 $g^{(m+1)} = \textrm{MM-GMOTV}(g^{(m)},{\bf S}^{(m)},\epsilon_q,\epsilon_m)$,
 with appropriately chosen termination tolerances $\epsilon_q$, and $\epsilon_m$.
 
Next, for solving step 2, we can use  MM algorithm developed in Section 3.2.
Specifically, we call MM-KL with ${\bf S}^{(m)}$  as initialization
for the minimization variable ${\bf S}$,  and  with ${\bf v}(x) = ({\bf L}*g^{(m+1)})(x)$.
In other words, the result of step 2 can be obtained as 
${\bf S}^{(m+1)}=\textrm{MM-KL}( ({\bf L}*g^{(m+1)})(x), {\bf S}^{(m)}, \lambda_F, 
\epsilon_{kl})$, with appropriately chosen termination tolerance $\epsilon_{kl}$.

Note that,  the iterative methods  $\textrm{MM-GMOTV}$  and $\textrm{MM-KL}$
terminate based on the gradient norms.  In our experiments,  we observed a good
convergence of the overall algorithm with high quality restoration results by setting
the termination tolerance to be lower than $10^{-4}$.  However, we are  not aware of
any theoretical results for the convergence of the overall iteration, when the iteration
for sub-problems are terminated based on the gradient norms.   In the following
proposition, we provide alternative termination conditions for the sub-problem
that can be met with finite number of inner iterations.
\begin{prop}
The algorithm expressed by Step 1 and Step 2 in the equations 
(\ref{eq:bcgmin})  and  (\ref{eq:bcsmin}), converges to a local minimum 
 if the following
conditions are satisfied:
\begin{equation}
\label{eq:step1term}
\left|\left< g^{(m+1)}-g^{(m)}, 
\nabla_g {J}_F(g^{(m+1)},{\bf S}^{(m)}) \right>\right|
< 
\left|\left< g^{(m+1)}-g^{(m)}, 
\nabla_g {J}_F(g^{(m)},{\bf S}^{(m)}) \right>\right|    
\end{equation}
\begin{align}
\label{eq:step2term}
\left|\left< {\bf S}^{(m+1)}-{\bf S}^{(m)}, 
 \nabla_{\bf S} {J}_F(g^{(m+1)},{\bf S}^{(m+1)}) \right>\right| 
  < 
 \left|\left< {\bf S}^{(m+1)}-{\bf S}^{(m)}, 
 \nabla_{\bf S} {J}_F(g^{(m+1)},{\bf S}^{(m)}) \right>\right|
\end{align}
\end{prop}

\section{Experimental results}
	
To evaluate the effectiveness of the proposed regularization, GMO-TV, we used ECG and EEG signals from
 MIT-BIH databases \cite{MIT-BIH}.   We tested the effectiveness  of the proposed GMO-TV approach in
 two variations:  in the first form,  we used first and second order derivatives, and in the second form, we used first to 
 fourth order  derivatives.  We denote the first form by GMO-TV2 and the second form by GMO-TV4. The 
 derivatives were implemented using the  discrete filters  [1 -1],  [1 -2 1],  [-1 3 -3 1] and  [1 -4 6 -4 1]. The choice of fourth order as the highest order derivative is  ad-hoc and the limit is allowable complexity. 
  The corresponding iterative versions, where training is eliminated,  is  referred to as IGMO-TV2 and
   IGMO-TV4. We compared our proposed approaches against the state of art TGV \cite{TGV2} and the recent 
   GTV \cite{GTV}. We also compared with the classic TV versions, denoted as TV1 and TV2.  To demonstrate the
   importance of combining higher order derivative with lower ones,  we also implemented two other variations:
   third and fourth order total variations, denoted by TV3, and TV4.
      To measure the restoration performance in our experiments, we used ISNR defined as 
\begin{equation}
\mathrm{ISNR} =20 \log_{10}\left(\frac{||g-f||_{2}}{||g-\hat{g}||_{2}}\right)  
\end{equation}
where $\hat{g}$ is the restored image, $f$ is the distorted input image and $g$ is the original image. For all cases, the tuning parameters including $\lambda$ were selected for each method to get the highest ISNR. The number of iterations was set large enough for TGV and GTV to ensure convergence in all experiments. In the case of GMO-TV2 and GMO-TV4 formulations with training samples, the stopping condition for MM iterations was gradient norm falling below $10^{-6}$. 
 
For implementing the training-free version,  the termination conditions for the
sub-problem of step 1 and step 2
(equations (\ref{eq:bcgmin}) and (\ref{eq:bcsmin}) were set  as 
$\left\|\nabla_g {J}_F(g^{(m+1)},{\bf S}^{(m)}) \right\|_2 < \epsilon_a$ and 
$\left\| \nabla_{\bf S} {J}_F(g^{(m+1)},{\bf S}^{(m+1)})\right\|_2 < \epsilon_a$
which we found to yield better results compared to conditions given in the
Proposition 3. For all experiments, $\epsilon_a$ was set to $10^{-6}$. Note that the
conditions we used are much stronger than the ones given in Proposition 3.

In the first experiment,  we considered  denoising of  ECG signals corrupted by 
additive white gaussian noise (AWGN) with the noise variances adjusted to match the listed  SNR values. The 
comparison results are given in Table \ref{Table_denoise_ecg_n}.  For the training mode,  the structure
 matrix {\bf S} was generated using  samples of ECG record 16272 from the MIT-BIH Normal Sinus Rhythm 
 database \cite{mitbih_nsrdb}.  For generating test signal, 2048 samples from ECG record 16265 was taken
 and divided into four segments,  each of 512 samples. Each experiment was performed on all four segments 
 and the results averaged to  get accurate performance results for all algorithms. The results show that GMO-TV4 
 gives the best performance in all cases followed closely by GMO-TV2,  IGMO-TV4 and IGMO-TV2.
 It should be emphasized that the training-free versions,  GMO-TV4 and IGMO-TV2 are clearly superior to
 TGV, and GTV. They are also superior to the single order total variations,  TV1---TV4.
 \begin{table}[!ht]`
		\centering
		\caption{Denoising Normal Sinus Rhythm ECG Signal}
		\label{Table_denoise_ecg_n}
		\resizebox{\textwidth}{!}
		{%
			\begin{tabular}{|c|c|c|c|c|c|c|c|c|c|c|}
				\hline
				SNR & TV1 & TV2 & TV3 & TV4 & GMO-TV4 & GMO-TV2 & IGMO-TV4 & IGMO-TV2 & TGV & GTV \\ \hline
				25 & 3.43 & 3.18 & 2.99 & 2.52 & {\bf 4.06} & 3.97 & 3.83 & 3.98 & 3.20 & 3.36 \\ \hline 
				20 & 4.06 & 3.75 & 3.68 & 3.25 & {\bf 4.86} & 4.64 & 4.82 & 4.67 & 3.74 & 4.04 \\ \hline 
				15 & 4.90 & 4.30 & 4.46 & 3.94 & {\bf 5.92} & 5.45 & 5.80 & 5.43 & 4.37 & 4.88 \\ \hline
				10 & 6.54 & 5.83 & 5.69 & 5.29 & {\bf 7.80} & 7.30 & 7.46 & 7.19 & 5.92 & 6.46 \\ \hline 
			\end{tabular}%
		}
\end{table}

In second experiment, we consider the deblurring problem.  We tested both training-based and training-free methods.
For generating test measurements,  we consider the same set of ECG signals used for the first experiment,
along with a new set of EEG signals.  We  used EEG record chb01\_02\_edfm from CHB-MIT Scalp EEG database \cite{chbmit_eeg} for training and four 512 length  segments from chb01\_01\_edfm record for  testing.
We considered four levels of Gaussian blurring by setting the variance of blurring kernel, $\sigma_{b}^{2}$  appropriately.
Also for each blurring level,  we considered four levels of AWGN noise.  The noise levels were chosen
such that the corresponding BSNR attains dB values $\{10,15,20,25\}$  where BSNR is defined as follows
 \cite{Unser_Hess13}:
	\begin{equation}
	\mathrm{BSNR} = \mathrm{var}(h \ast g)/\sigma_{\eta}^{2}
	\end{equation}
	The results for ECG  and EEG test signals are presented in the  Table \ref{Table_deblur_ecg_n} and Table \ref{Table_deblur_eeg_scalp} respectively. 
\begin{table}[!ht]
		\centering
		\caption{Deblurring Normal Sinus Rhythm ECG Signal}
		\label{Table_deblur_ecg_n}
		\resizebox{\textwidth}{!}
		{%
			\begin{tabular}{|c|c|c|c|c|c|c|c|c|c|c|}
				\hline
				BSNR & $\sigma^{2}_{b}$ & TV1 & TV2 & TV3 & TV4 & GMO-TV4 & GMO-TV2 & IGMO-TV4 & IGMO-TV2 & TGV \\ \hline
				\multirow{4}{*}{25} & 1 & 5.73 & 5.88 & 8.76 & 7.95 & 9.61 & 9.95 & 8.96 & {\bf 9.97} & 7.53 \\ \cline{2-11} 
				& 2 & 5.97 & 7.22 & 10.42 & 9.76 & 10.98 & {\bf 12.11} & 10.99 & 11.28 & 9.45 \\ \cline{2-11} 
				& 4 & 7.72 & 9.27 & 11.39 & 10.79 & 12.81 & {\bf 13.57} & 11.59 & 12.62 & 8.68\\ \cline{2-11} 
				& 6 & 7.65 & 9.28 & 9.87 & 8.82 & 12.05 & 12.38 & 10.04 & {\bf 12.61} & 6.59 \\ \hline
				\multirow{4}{*}{20} & 1 & 2.69 & 4.20 & 7.30 & 6.33 & 7.85 & {\bf 8.09} & 7.37 & 8.07 & 4.75 \\ \cline{2-11} 
				& 2 & 5.25 & 6.88 & 8.92 & 8.85 & 9.76 & {\bf 10.42} & 9.22 & 9.85 & 7.06 \\ \cline{2-11} 
				& 4 & 6.31 & 8.14 & 9.13 & 8.52 & 10.80 & 11.06 & 9.90 & {\bf 11.13}  & 6.64 \\ \cline{2-11} 
				& 6 & 6.97 & 8.19 & 8.45 & 7.66 & 10.95 & {\bf 10.96} & 7.35 & 9.83 & 5.71 \\ \hline
				\multirow{4}{*}{15} & 1 & 2.65 & 3.54 & 5.40 & 4.90 & 6.20 & {\bf 6.46} & 4.17 & 4.28 & 3.11 \\ \cline{2-11} 
				& 2 & 4.46 & 5.71 & 7.16 & 6.86 & 8.30 & 8.46 &7.66 & {\bf 8.51} & 4.86 \\ \cline{2-11} 
				& 4 & 5.35 & 6.38 & 6.57 & 6.06 & {\bf 8.48} & 8.32 & 7.81 & 8.28 & 5.29 \\ \cline{2-11} 
				& 6 & 5.45 & 6.18 & 5.97 & 5.50 & {\bf 8.47} & 8.08 & 6.49 & 7.82 & 3.76 \\ \hline
				\multirow{4}{*}{10} & 1 & 3.00 & 3.62 & 4.02 & 3.62 & {\bf 5.53} & 5.34 & 1.71 & 3.37 & 2.30 \\ \cline{2-11} 
				& 2 & 3.52 & 4.42 & 4.52 & 3.85 & {\bf 6.37} & 6.30 & 3.86 & 6.00 & 3.01 \\ \cline{2-11} 
				& 4 & 4.34 & 4.79 & 4.70 & 4.10 & {\bf 6.77} & 6.44 & 6.46 & 5.64 & 3.65 \\ \cline{2-11} 
				& 6 & 3.82 & 4.02 & 3.83 & 3.35 & {\bf 6.12} & 5.53 & 6.61 & 5.26 & 2.55 \\ \hline
			\end{tabular}%
		}
	\end{table}
\begin{table}[!ht]
		\centering
		\caption{Deblurring Scalp EEG Signal}
		\label{Table_deblur_eeg_scalp}
		\resizebox{ \textwidth}{!}{%
			\begin{tabular}{|c|c|c|c|c|c|c|c|c|c|c|}
				\hline
				BSNR & $\sigma^{2}_{b}$ & TV1 & TV2 & TV3 & TV4 & GMO-TV4 & GMO-TV2 & IGMO-TV4 & IGMO-TV2 & TGV \\ \hline
				\multirow{4}{*}{25} & 1 & 1.48 & 2.37 & {\bf 2.82} & 2.49 & 2.81 & 2.77 & 2.34 & 2.42 & 2.58 \\ \cline{2-11}
				& 2 & 1.10 & 2.74 & 3.06 & 3.07 & {\bf 3.12} & 2.84 & 2.72 & 2.81 & 2.95 \\ \cline{2-11}
				& 4 & 1.55 & 3.05 & 3.32 & 3.33 & 3.46 & 3.24 & {\bf 3.50} & 3.38 & 3.47 \\ \cline{2-11}
				& 6 & 1.25 & 2.61 & 2.72 & 2.81 & 2.77 & 2.64 & {\bf 3.31} & 3.08 & 2.83 \\ \hline
				\multirow{4}{*}{20} & 1 & 0.29 & 1.78 & 1.73 & 1.8 & {\bf 1.99} & 1.82 & 1.18 & 1.52 & 1.51 \\ \cline{2-11}
				& 2 & 0.72 & 2.23 & 2.33 & 2.41 & {\bf 2.44} & 2.21 & 1.94 & 2.08 & 2.35 \\ \cline{2-11}
				& 4 & 0.96 & 2.21 & 2.40 & 2.42 & 2.48 & 2.29 & {\bf 2.98} & 2.81 & 2.78 \\ \cline{2-11}
				& 6 & 1.29 & 2.52 & 2.68 & {\bf 2.66} & 2.59 & 2.43 & 2.27 & 2.27 & 2.21 \\ \hline
				\multirow{4}{*}{15} & 1 & 0.57 & 1.41 & 1.45 & 1.47 & {\bf 1.66} & 1.57 & 0.32 & 1.25 & 1.43 \\ \cline{2-11}
				& 2 & 1.08 & 2.16 & 2.36 & 2.42 & {\bf 2.49} & 2.31 & 1.71 & 2.15 & 2.00 \\ \cline{2-11}
				& 4 & 1.46 & 2.47 & 2.61 & 2.62 & {\bf 2.63} & 2.50 & 2.31 & 2.24 & 2.33 \\ \cline{2-11}
				& 6 & 1.27 & 2.13 & 2.18 & 2.14 & {\bf 2.29} & 2.21 & 2.20 & 2.25 & 2.27 \\ \hline
				\multirow{4}{*}{10} & 1 & 1.73	& 2.54 & 2.59 & 2.57 & {\bf 2.76} & 2.58 & 0.95 & 1.78 & 1.58 \\ \cline{2-11}
				& 2 & 1.70 & 2.46 & 2.52 & 2.47 & 2.72 & 2.56 & 1.68 & 2.20 & {\bf 2.88} \\ \cline{2-11}
				& 4 & 1.99 & 2.52 & 2.66 & 2.59 & 2.68 & 2.63 & 2.63 & {\bf 2.83} & 2.59 \\ \cline{2-11}
				& 6 & 1.80 & 2.46 & 2.53 & 2.49 & 2.49 & 2.42 & 2.52 & {\bf 2.68} & 2.50 \\ \hline
			\end{tabular}%
		}
	\end{table}

\begin{figure}[!ht]
		\centering
		\includegraphics[width = \textwidth]{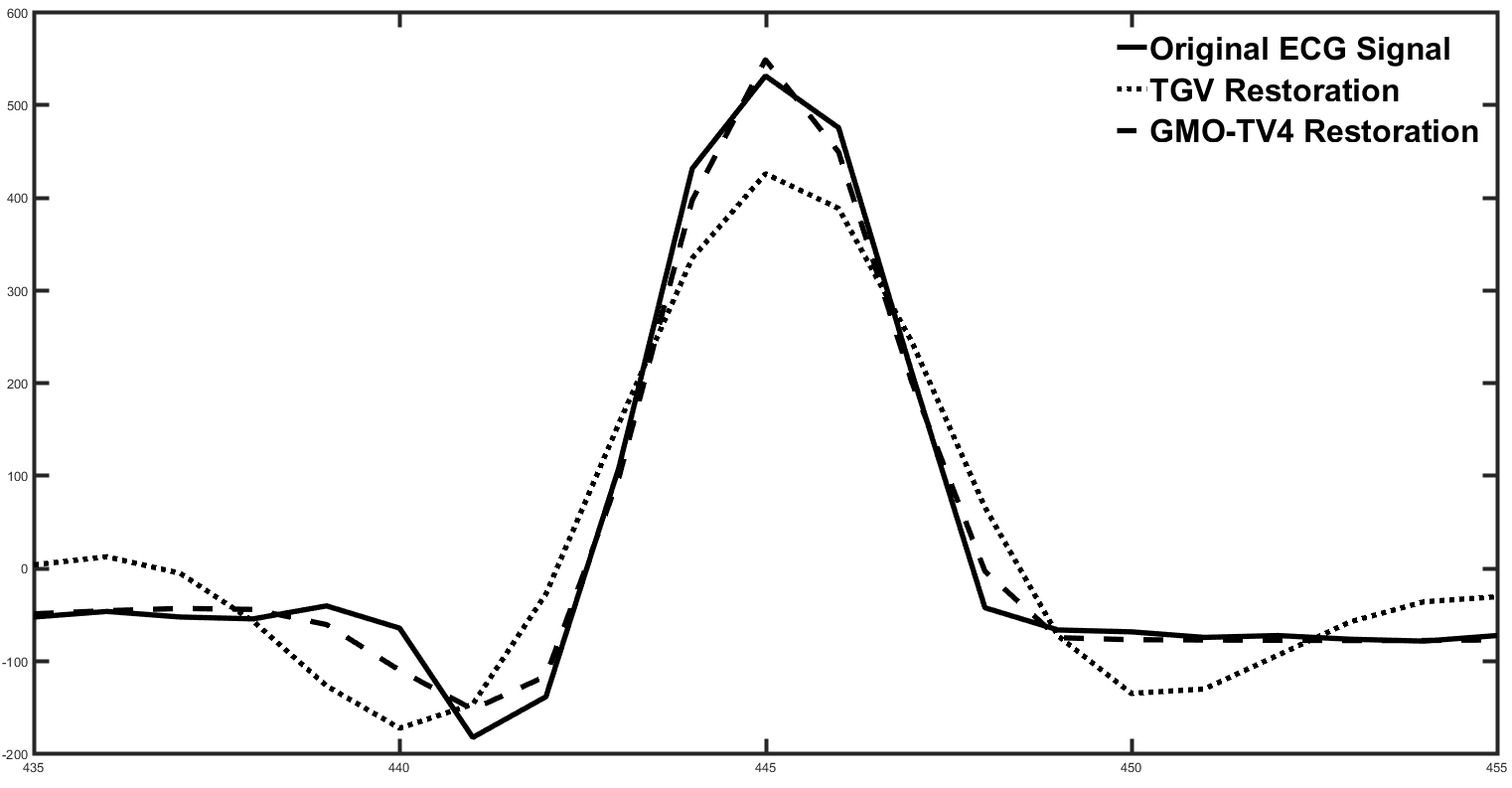}
		\begin{center}
			\caption{Deconvolution of ECG Signal (BSNR=25, $\sigma_b^{2}$=4)}
		\end{center}
		\label{Fig_ECG_Sel}
	\end{figure}

The results offer some interesting insights into the working of proposed approaches. While GMO-TV2 and IGMO-TV2 
utilizing first and second order derivatives perform better at high BSNR values for ECG restoration, GMO-TV4 and 
IGMO-TV4 perform better at lower BSNR values. This indicates that higher order derivatives are robust to noise, as seen in the denoising experiment.
Similarly at high BSNR values, IGMO-TV2 is able to give performance comparable to learning based GMO-TV4 and 
GMO-TV2 or even better in some cases. This indicates that training from noise-free samples helps in increased performance only for measurements at high noise levels. In other cases, the measurements themselves are sufficient for building the structure matrix $\S$. 
 Furthermore in all cases, IGMO-TV2 and IGMO-TV4 perform  better than other TV based functionals including TGV, 
 and GTV, demonstrating the power of  our formulation even without any training samples. Figure \ref{Fig_ECG_Sel} shows the restoration result with ECG signal corresponding to BSNR =25 and 
 $\sigma_b^{2}$=4. 

While ECG signals have a  discernible structure and the restoration results from Table \ref{Table_deblur_ecg_n} indicate that the proposed approaches can utilize the same with or without training samples, exploiting the signal structure in
EEG signals is much more challenging. The restoration results in Table \ref{Table_deblur_eeg_scalp} show that the proposed approaches give better performance than higher order TV as well as TGV in most cases. But unlike the case of ECG, where the proposed approaches gave around 2-4dB improvement over other TV functionals, the difference in ISNR between the techniques is around 0.1-1dB in the case of EEG.  This is because of the fact that EEG signals are not as structured as ECG signals.
Besides,  TGV has an advantage that it is spatially adaptive because of the   auxiliary variable involved in its definition. 
Nevertheless, all four variants of the proposed method  including the training-free ones perform better than TGV
in most cases. 
	
	\section{Conclusion}
	
We  proposed a novel total variation based regularization functional  named
Generalized Multi-Order Total Variation (GMO-TV) that exploits dependencies
 among multiple order signal derivatives. We derived the functional from 
 cross-entropy formulation by adopting a form of multivariate Laplacian prior 
 probability for multiple order signal derivatives. The new prior allows the 
 regularization functional to be adaptive to the patterns of intensity variations 
 that are specific to the class of signals under consideration. The adaptivity is 
 achieved by the means of a structure matrix either  built via training, or
 estimated jointly along with the required signal via minimization. 
  We demonstrated, using experimental examples,  that GMO-TV outperforms 
 standard TVs as well as TGV and GTV functionals with or without training.

	
\section*{Appendix}

\subsection*{Proof of proposition 1}

Let ${\bf p} =  [{\bf p}_1^T\; {\bf p}_2^T\; \dots {\bf p}_K^T]^T$, where ${\bf p}_i$'s are
vectors of size $K\times 1$ such that 
${\bf S} = 
\left[{\bf p}_1\; {\bf p}_2\; \ldots \; {\bf p}_K\right]^T$. 
For notational convenience in deriving the
algorithm, we re-express ${R}_F(g, {\bf S})$ in terms of ${\bf p}$ and ${\bf v}(x)$
as given below:
\begin{align}
\label{eq:kpv}
Z({\bf p}, {\bf v}) = 
 \sum_{x} \sqrt{\sum_{i=1}^K({\bf p}_i^T{\bf v}(x))^2} &+ \frac{1}{2}\lambda_F \|{\bf p}\|_2^2 \\ \nonumber&- \frac{1}{2}\log(\det| \left[{\bf p}_1\; {\bf p}_2\; \ldots \; {\bf p}_K\right]^T\left[{\bf p}_1\; {\bf p}_2\; \ldots \; {\bf p}_K\right]|) 
 \end{align}
Since the  vectors ${\bf p}_i$'s are orthogonal, we get 
\begin{equation}
\label{eq:kpv2}
Z({\bf p}, {\bf v}) =  
 \sum_{x} \sqrt{\sum_{i=1}^K({\bf p}_i^T{\bf v}(x))^2} - 
 \frac{1}{2}\log(\prod_{i=1}^{K} ||\mathbf{p}_{i}||_{2}^{2}) + \frac{1}{2}\lambda_F
 \sum_{i=1}^{K} ||\mathbf{p}_{i}||_{2}^{2}
 \end{equation}
 Taking gradient of $Z({\bf p}, {\bf v})$  with respect to each ${\bf p}_i$ gives
 \begin{equation}
 \label{eq:gradkpv}
\nabla_{{\bf p}_i} Z({\bf p}, {\bf v}) = \sum_x \frac{1}
{\sqrt{\sum_{j=1}^K({\bf p}_j^T{\bf v}(x))^2}}
{\bf v}(x){\bf v}^{T}(x)\mathbf{p}_i
- \frac{{\bf p}_i}{\|\mathbf{p}_{i}\|_{2}^{2}} + \lambda_F \mathbf{p}_{i}
\end{equation}
Let  $\nabla_{\bf S}\bar{R}(g, {\bf S})$ denote gradient with respect to whole ${\bf S}$ in matrix form.
Then 
 \begin{equation}
 \label{eq:gradkpv2}
 \nabla_{\bf S}\bar{R}(g, {\bf S}) = \left[ \nabla_{{\bf p}_1} Z({\bf p}, {\bf v})  \ldots \nabla_{{\bf p}_K} Z({\bf p}, {\bf v}) \right]^T
 \end{equation}
 Combining the equations  (\ref{eq:gradkpv}) and  (\ref{eq:gradkpv2}) gives
 $\nabla_{\bf S}\bar{R}(g,{\bf S})  = {\bf S}{\bf A} - 
 ({\bf S}{\bf S}^T)^{-1}{\bf S} + \lambda_F {\bf S}$, where
 ${\bf A}$ = ${\sum \limits_{x}} \frac{1}{\|{\bf S}(({\bf L}*g)(x))\|_2}(({\bf L}*g)(x))(({\bf L}*g)(x))^T$.
 Using similar steps,   we get gradient for $\bar{R}^{(k)}(g,{\bf S},{\bf S}^{(k)})$
 as given below:
  \begin{equation}
\label{eq:gradjsm_proof}
\nabla_{\bf S}\bar{R}^{(k)}(g,{\bf S},{\bf S}^{(k)})  =  {\bf S}{\bf A}_k - 
 ({\bf S}{\bf S}^T)^{-1}{\bf S}+ \lambda_F {\bf S},
\end{equation}
where ${\bf A}_{k} = {\sum \limits_{x}} \frac{1}{\|{\bf S}^{(k)}({\bf L}*g)(x))\|_2}({\bf L}*g)(x))({\bf L}*g)(x))^T$
 
 \subsection*{Proof of proposition 2}
 
 To minimize $\bar{R}^{(k)}(g,{\bf S},{\bf S}^{(k)})$, we equate the gradient to zero:
 $${\bf S}{\bf A}_k - 
 ({\bf S}{\bf S}^T   )^{-1}{\bf S} + \lambda_F {\bf S}= {\bf 0}$$
 Using the fact that ${\bf A}_k$ is symmetric,  re-write the above equation for individual
 rows of ${\bf S}$  as given below:
 \begin{equation}
\mathbf{A}_{k}\mathbf{p}_{i} - \frac{1}{||\mathbf{p}_{i}||_{2}^{2}}\mathbf{p}_{i} 
+ \lambda_F \mathbf{p}_{i}= 0, \; i =1,\ldots,K.
\end{equation}
The above equation implies that $\mathbf{p}_{i}$  are of the form  
$\mathbf{p}_{i} = \beta_i{\bf e}_i$ where ${\bf e}_i$ is the $i$th Eigen vector of
$\mathbf{A}_{k}$ and $\beta_i$ is non-negative factor.  Substituting this in the above 
equation gives
\begin{equation}
\label{eq:eigeq}
 \beta_{i}\eta_{i} \mathbf{e}_{i} - \frac{1}{\beta_{i}^{2}}\beta_{i} \mathbf{e}_{i}
 + \lambda_F \beta_{i} \mathbf{e}_{i}
= 0, \;\;
i=1,\ldots,K.
\end{equation}
where $\eta_{i}$ is the corresponding Eigen value.  The above equation 
 gives  $\beta_{i} = \frac{1}{\sqrt{\eta_{i}+\lambda_F}}$. 
 This means that if ${\bf S}^{(k+1)}$ be the minimum of
${R}^{(k)}_F(g,{\bf S},{\bf S}^{(k)})$,   then ${\bf S}^{(k+1)}  =
( {\bf D}_k+\lambda_F)^{-1/2} {\bf U}_k^T$ where 
${\bf D}_k$  and ${\bf U}_k$ are the matrices involved in the Eigen decomposition of
${\bf A}_{k}$, i.e., ${\bf A}_{k}  = {\bf U}_k{\bf D}_k {\bf U}_k^T$.

\subsection*{Proof of proposition 3}

Let ${\bf p}$ denote the rows of   ${\bf S}$ stacked vertically and let ${\bf g}$
denote the samples of $g$ in vector form.  Let ${\bf y} = [{\bf g}^T {\bf p}^T]^T$.
Let $J_v({\bf y})$ is the function defined on ${\bf y}$   such that 
$J_v({\bf y} )=J_F(g,{\bf S})$.    Let ${\bf g}_m$ denote the vector corresponding
to $g^{(m)}$ and let ${\bf p}_m$ denote the vector corresponding to
 ${\bf S}^{(m)}$.   Let ${\bf y}_{2m-1} = [{\bf g}^T_{m} \;\;{\bf p}^T_{m-1}]^T$  and
 ${\bf y}_{2m} = [{\bf g}^T_{m}\;\; {\bf p}^T_{m}]^T$.  Then ${\bf y}_{l=1,2,\ldots}$
 denotes the sequence of iterates generated by the algorithm.  Note that
 ${\bf d}_{l-1} = {\bf y}_l - {\bf y}_{l-1}$ is the  search direction at the point ${\bf y}_{l-1}$.
Note that for odd  $l$, ${\bf d}_l$ is non-zero only for the variable ${\bf p}$;
similarly,  for even $l$, ${\bf d}_l$ is non-zero only for the variable ${\bf g}$.
Hence for any $l$,  ${\bf d}_l$ is non-zero for parts corresponding to only one
of the variables in $\{{\bf g}, {\bf p}\}$.  Further,  note that the function
 $J_v({\bf y})$ is convex with respect to any  one of the variables in $\{{\bf g}, {\bf p}\}$.
 The last two statements imply that all ${\bf d}_l$'s  are  descent directions.
 For each descent direction ${\bf d}_l$, the update ${\bf y}_l = {\bf y}_{l-1} + {\bf d}_{l-1}$
 can be considered as a result of line search.  
 
 Then according to Zoutendijk Lemma \cite{nocedal2006},      such a series of line searches
 along descent directions converge to a local minimum, if the following conditions
 are satisfied:  (1) the sub-level set  of $J_v({\bf y})$ for initialization ${\bf y}_0$
 is bounded; (2) the gradient of $J_v({\bf y})$  is Lipschitz continuous;  
 (3) the line search satisfies Wolfe's condition, i.e.,  
 \begin{equation}
 \label{eq:wolfe}
 \left|{\bf d}_{l}^T\nabla_{\bf y} J_v({\bf y}_l) \right|
  <  \left|{\bf d}_{l}^T\nabla_{\bf y} J_v({\bf y}_{l-1}) \right|
  \end{equation}
  
  Since the function $J_v({\bf y})$ is bounded below,  the first condition is satisfied.
  Also,  the gradient is obviously Lipschitz continuous.  Now, note that the condition
  given in the equation (\ref{eq:wolfe}),  can be written in two forms for odd and
  even values of $l$ as given below:
  \begin{align}
  \label{eq:wolfe2}
  \left|({\bf g}_m - {\bf g}_{m-1})^T\nabla_{\bf g}J_v\left(\left[
{\bf g}_{m}^T \;\; {\bf p}_{m-1}^T  \right]^T\right)\right| & < 
\left|({\bf g}_m - {\bf g}_{m-1})^T\nabla_{\bf g}J_v\left(\left[
{\bf g}_{m-1}^T \;\; {\bf p}_{m-1}^T  \right]^T\right)\right|  \\
 \left|({\bf p}_m - {\bf p}_{m-1})^T\nabla_{\bf g}J_v\left(\left[
{\bf g}_{m}^T \;\; {\bf p}_{m}^T  \right]^T\right)\right| & < 
\left|({\bf p}_m - {\bf p}_{m-1})^T\nabla_{\bf g}J_v\left(\left[
{\bf g}_{m}^T \;\; {\bf p}_{m-1}^T  \right]^T\right)\right|
  \end{align}
 Rewriting the above equations in terms of the original variables $g$ and ${\bf S}$
 by taking into account the dependence of the sub-parts of $J_F(g,{\bf S})$ on
 the variables $g$ and ${\bf S}$,  we get the conditions of the Proposition 3.

\bibliographystyle{unsrt}  
\bibliography{GMOTV}

\begin{thebibliography}{10}

\bibitem{Rudin_tv1}
Leonid~I. Rudin, Stanley Osher, and Emad Fatemi.
\newblock Nonlinear total variation based noise removal algorithms.
\newblock {\em Physica D: Nonlinear Phenomena}, 60(1–4):259 -- 268, 1992.

\bibitem{TV_ECGSmooth}
Om~Prakash Yadav and Shashwati Ray.
\newblock Smoothening and segmentation of {ECG} signals using total variation
  denoising-minimization-majorization and bottom-up approach.
\newblock {\em Procedia Computer Science}, 85:483 -- 489, 2016.

\bibitem{TV_FetalECG}
Kwang~Jin Lee and Boreom Lee.
\newblock Sequential total variation denoising for the extraction of fetal
  {ECG} from single-channel maternal abdominal {ECG}.
\newblock {\em Sensors}, 16(7):1020, 2016.

\bibitem{TV_Chambolle04}
Antonin Chambolle.
\newblock An algorithm for total variation minimization and applications.
\newblock {\em Journal of Mathematical Imaging and Vision}, 20(1):89--97, 2004.

\bibitem{TV_Chan_BlindDecon}
T.~F. Chan and Chiu-Kwong Wong.
\newblock Total variation blind deconvolution.
\newblock {\em IEEE Transactions on Image Processing}, 7(3):370--375, 1998.

\bibitem{TV_CMDecon}
N.~Dey, L.~Blanc-Feraud, C.~Zimmer, Z.~Kam, J.~C. Olivo-Marin, and J.~Zerubia.
\newblock A deconvolution method for confocal microscopy with total variation
  regularization.
\newblock In {\em IEEE International Symposium on Biomedical Imaging: Nano to
  Macro}, pages 1223--1226 Vol. 2, 2004.

\bibitem{TV_LiBlindDecon}
Weihong Li, Quanli Li, Weiguo Gong, and Shu Tang.
\newblock Total variation blind deconvolution employing split bregman
  iteration.
\newblock {\em Journal of Visual Communication and Image Representation},
  23(3):409 -- 417, 2012.

\bibitem{tikh_1}
A.~N. Tikhonov and V.~Y. Arsenin.
\newblock Solution of ill-posed problems.
\newblock {\em V.H. Winston, Washington, DC}, 1977.

\bibitem{TV_GpuPock}
T.~Pock, M.~Unger, D.~Cremers, and H.~Bischof.
\newblock Fast and exact solution of total variation models on the {GPU}.
\newblock In {\em IEEE Computer Society Conference on Computer Vision and
  Pattern Recognition Workshops}, pages 1--8, 2008.

\bibitem{TV_GpuCT}
Xun Jia, Yifei Lou, Ruijiang Li, William~Y. Song, and Steve~B. Jiang.
\newblock {GPU}-based fast cone beam {CT} reconstruction from undersampled and
  noisy projection data via total variation.
\newblock {\em Medical Physics}, 37(4):1757--1760, 2010.

\bibitem{TV_Ring}
Wolfgang Ring.
\newblock Structural properties of solutions to total variation regularization
  problems.
\newblock {\em ESAIM: Mathematical Modelling and Numerical Analysis},
  34:799--810, 2000.

\bibitem{Combined_order_TV}
K.~Papafitsoros and C.~B. Sch{\"o}nlieb.
\newblock A combined first and second order variational approach for image
  reconstruction.
\newblock {\em Journal of Mathematical Imaging and Vision}, 48(2):308--338,
  2014.

\bibitem{Scherzer_tv2_98}
O.~Scherzer.
\newblock Denoising with higher order derivatives of bounded variation and an
  application to parameter estimation.
\newblock {\em Computing}, 60(1):1--27, 1998.

\bibitem{TV_CombinedLysaker}
Marius Lysaker and Xue-Cheng Tai.
\newblock Iterative image restoration combining total variation minimization
  and a second-order functional.
\newblock {\em International Journal of Computer Vision}, 66(1):5--18, 2006.

\bibitem{Bergounioux2010}
Ma{\"i}tine Bergounioux and Loic Piffet.
\newblock A second-order model for image denoising.
\newblock {\em Set-Valued and Variational Analysis}, 18(3):277--306, Dec 2010.

\bibitem{Bredies_TGV_2010}
Kristian Bredies, Karl Kunisch, and Thomas Pock.
\newblock Total generalized variation.
\newblock {\em SIAM J. Img. Sci.}, 3(3):492--526, 2010.

\bibitem{TGV2}
Florian Knoll, Kristian Bredies, Thomas Pock, and Rudolf Stollberger.
\newblock Second order total generalized variation ({TGV}) for {MRI}.
\newblock {\em Magnetic Resonance in Medicine}, 65(2):480--491, 2011.

\bibitem{GTV}
I.~W. Selesnick.
\newblock Generalized total variation: Tying the knots.
\newblock {\em IEEE Signal Processing Letters}, 22(11):2009--2013, Nov 2015.

\bibitem{NNCG_Deepak}
D.~G. Skariah and M.~Arigovindan.
\newblock Nested conjugate gradient algorithm with nested preconditioning for
  non-linear image restoration.
\newblock {\em IEEE Transactions on Image Processing}, 26(9):4471--4482, Sept
  2017.

\bibitem{Bertsekas_NLP}
Dimitri~P Bertsekas.
\newblock {\em Nonlinear programming}.
\newblock Athena scientific Belmont, 1999.

\bibitem{MIT-BIH}
Ary~L. Goldberger, Luis A.~N. Amaral, Leon Glass, Jeffrey~M. Hausdorff,
  Plamen~Ch. Ivanov, Roger~G. Mark, Joseph~E. Mietus, George~B. Moody,
  Chung-Kang Peng, and H.~Eugene Stanley.
\newblock Physiobank, physiotoolkit, and physionet.
\newblock {\em Circulation}, 101(23):e215--e220, 2000.

\bibitem{mitbih_nsrdb}
Mit-bih nsrdb, https://www.physionet.org/physiobank/database/nsrdb/.

\bibitem{chbmit_eeg}
Chb-mit db, https://www.physionet.org/physiobank/database/chbmit/.

\bibitem{Unser_Hess13}
S.~Lefkimmiatis, J.~P. Ward, and M.~Unser.
\newblock Hessian schatten-norm regularization for linear inverse problems.
\newblock {\em IEEE Transactions on Image Processing}, 22(5):1873--1888, 2013.

\bibitem{nocedal2006}
J.~Nocedal and S.~Wright.
\newblock {\em Numerical Optimization}.
\newblock Springer Series in Operations Research and Financial Engineering.
  Springer New York, 2006.

\end{thebibliography}

\end{document}